\documentclass[twocolumn,a4paper,prb,showpacs]{revtex4}
\usepackage{amssymb,amsfonts,amsmath,graphicx,bm}
\begin{document}

\title{Surface electronic structure for Al(111) by a scattering method}
\date{}
\author{M. N. Read}

\affiliation{School of Physics, University of New South Wales, \\ 
Sydney NSW
2052, Australia}
\email{mnr@phys.unsw.edu.au}

\begin{abstract}
A scattering method is used to calculate the surface band structure of Al(111) from 8.6 eV below the Fermi level to 9 eV above it. This method has rarely been implemented previously. The complete complex bulk and surface band structure is also shown for key values of the crystal surface-parallel momentum. The surface region is treated semi-empirically from a small number of experimentally determined quantities. We find agreement over the whole surface Brillouin zone for surface states compared with recent theoretical \textit{ab initio} results and experiment. Differences with respect to surface resonances detected by recent theoretical methods are discussed. In particular a class of surface resonances is detected here that may be surface-layer localised and results from very wide partial surface-projected energy gaps where real lines in the complex bulk band structure extend from band minima to infinity. Other methods do not appear to have detected these resonances. They may have an effect on photoemission intensity profiles and may account for previous unexplained experiment features considered to be ``a kind of surface enhancement at partial gap edges". This scattering method provides a single picture for the formation and classification of surface states and resonances. 
\end{abstract}
\pacs{73.20.At, 68.37.Nq, 61.05.jh}
\maketitle

\section{Introduction}

We use the scattering matrix method of McRae \cite{mcrae1} to calculate the surface band structure $E(\bm{k}_{\parallel})$ for semi-infinite Al(111) to 9 eV above the Fermi energy for $\overline{M}$ -- $\overline{K}$ -- $\overline{\Gamma}$ -- $\overline{M}$ of the surface Brillouin zone and show the complete near-surface energy band structure $E(\bm{k}_{\parallel}, k_{\perp})$ for symmetry points $\overline{\Gamma}$, $\overline{M}$, $\overline{K}$ and some points along the direction $\overline{\Gamma}$ -- $\overline{M}$$(\overline{\Sigma})$. These complete $E(\bm{k}_{\parallel}, k_{\perp})$ band structures show bulk states with propagating and attenuating wave functions as well as all surface states and resonances (SS/SRs) that are needed in the interpretation of photoemission and other surface sensitive phenomena. Reliable methods for clean metal surfaces are needed before calculations for more complex systems of foreign-atom layers on metal substrates can be attempted.

There have been a number of calculations of band structure for this surface by various methods \cite{boudreaux,caruthers,chelikovsky,mednick,wang,heinrich1,benesh,heinrich2,hummel} and recent methods \cite{benesh,heinrich2,hummel} including the present one agree on the occurrence of surface states. All the SS/SRs that have been found experimentally to-date are found in the recent calculations and present one. However even the recent \textit{ab initio} self-consistent calculations with free-form surface potentials in Refs.~\onlinecite{benesh,heinrich2,hummel} do not give consistent results for SRs and this does not inspire confidence in the reliability of any of the calculations for this case. The surface-embedded Green-function (SEGF) \cite{benesh} method identifies SS/SRs from the calculation of the layer-density-of-states (LDOS). The repeated-slab density-functional (DF) method \cite{heinrich2} identifies the SS/SRs from maxima near the surface and decay into the bulk of the planar charge density as a function of the coordinate perpendicular to the surface and also from the LDOS. The Òassembly of boundary-controlled monolayersÓ (ABCM) \cite{hummel} method relies on matching attenuating crystal wave functions to vacuum wave functions. The present scattering method identifies SS/SRs from a condition for confinement of the electron in the surface region due to sustained multiple scattering in that region. Our result agrees very closely with that of ABCM for a particular class of surface resonances. In the present scattering method we also find another class of surface resonances that would not be detected in the ABCM with the search method used in that particular case. Some of these ``extra" resonances but not all, appear in the slab DF calculation. If reasons can be found for the different results between methods then the most reliable methods can be identified. Agreement between calculated surface band structures from different theoretical approaches would provide confidence in the results and aid in the analysis of measured features found in inverse and direct photoemission.

From our calculation we suggest that the resonances that are missing in the work of Ref.~\onlinecite{hummel} occur in surface-projected bulk-band gaps or partial gaps that are very wide in energy and contain ``real lines" with real energy in the complex band structure where the minima of bulk bands have no lower-energy band maxima to which to join and extend to infinity. This type of resonance is often very surface-layer localized and may have been missed in the search method of Ref.~\onlinecite{hummel}. Additionally it is suggested that unexplained features in experimental angle-resolved photoemission spectroscopy (ARPES) that have previously been termed ``partial bulk gap-edge features" arising from ``a kind of surface enhancement at partial gap edges" \cite{hofman} are surface resonances of this type as these resonances often occur at energies just outside other narrower absolute or partial projected bulk gap-edges that lie within the energy range of the very wide gap. Such resonances are just as easy to detect using the present method as are SSs. It is suggested that this property gives the method some advantage over other methods. 

The above recent calculations are \textit{ab initio} and determine the surface potential self-consistently from a free-form shape. In most cases they use a local-density approximation for the exchange-correlation potential and do not produce the image potential in the vacuum that varies as the inverse of distance from the surface. The present method determines the non-spherically symmetric surface potential semi-empirically and other key properties from experiment. The geometric structure of the surface layers is determined from the result of previous low-energy electron diffraction (LEED) analysis. The spatial variation of the valence charge density at the crystal-vacuum interface and hence the surface barrier potential is found from previously determined experimental energies of image barrier and other SS/SRs at $\overline{\Gamma}$ and the experimental work function. The spherically-symmetric bulk ion-core potentials and constant interstitial potential are determined from previous self-consistent bulk-band structure calculations that have shown agreement with experimental bulk states. Experimental surface core-level binding energy shifts for Al(111) are negligible \cite{borg} and final-state relaxation contributions are small for simple metals.\cite{riffe,alden} This leads to choosing the surface-layer constant potential and the surface ion-core potential the same as the bulk in this case. The non-spherically symmetric surface potential is then effectively an overlap of the surface barrier potential and spherically-symmetric surface ion-core potential. In this scattering formulation ion-core potentials are represented as point sources once their scattering phase shifts are calculated and their radius is no longer of relevance. It is to be noted that there are no discontinuities in the constant potential or its derivative in the surface region in these calculations that might give rise to spurious non-physical features.

While this surface potential may seem too crude an approximation in comparison with the \textit{ab initio} self-consistent surface potentials of the above recent calculations it adequately produces all SS/SRs found from those methods in this case. It can also be used at energies above the vacuum level where inelastic effects become significant. Its applicability to cases where surface core levels shifts are expected and for overlayers of foreign atoms at the surface is to be assessed in the future. It may have use for systems that are too complicated for the \textit{ab initio} methods. 

\section{Method of calculation}

A scattering method is used to determine the electronic properties. The semi-crystal is divided into atomic layers parallel to the surface. Between these layers are infinitesimally thin 2-dimensional (2D) periodic layers of constant potential where the wave function may be expanded in terms of a sufficient number of forward and backward travelling plane waves. A plane wave inside the crystal scatters from the top atomic layer with energy $E_1$ relative to the constant potential energy between ion-core potential wells in the layer, $E_{c,1}$ and with surface-parallel momentum or wave vector $\bm{k}_{\parallel}$. In the general case the bulk layers may have a different constant potential energy, $E_c$ and the energy with respect to $E_c$ is $E$. In the present case for Al(111) we have $E_{c,1}=E_c$ that will be justified later. The wave function of a layer is expanded into spherical waves and matched to plane waves (propagating and evanescent or attenuating in space) in the regions of constant potential between the layers using the layer-Korringa-Kohn-Rostocker (LKKR) Green function method of Kambe. \cite{kambe} Partial-wave phase shifts with angular momentum $l$ describe the ion-core scattering. It should be noted that rapid convergence is obtained in this calculation for these energies because sums over both direct and reciprocal net vectors are performed in the method of Kambe. The scattering properties of a layer are described by a layer transfer matrix that contains complex amplitude reflection and transmission coefficients. 

The method of McRae \cite{mcrae2} is used to combine layer scattering properties to obtain the scattering properties of the semi-infinite crystal and hence the electronic properties. For bulk layers that become the substrate of the semi-crystal, the layer transfer matrix is combined with the bulk geometrical structure to form the reduced transfer matrix $\bm{R}$ of the crystal substrate. From the eigenvectors of $\bm{R}$ the matrix of complex amplitude reflection coefficients for the whole semi-infinite crystal substrate $\bm{M}_1$ is obtained. Further transfer matrices for special surface atomic layers may be combined with $\bm{M}_1$ to form the matrix $\bm{M}$ that contains the complex amplitude reflection coefficients for scattering from all atomic layers of the semi-infinite crystal. The scattering properties of the top-most surface barrier layer of potential are contained in the matrix $\bm{S}$. The sub-matrix $\bm{S}^{\text{\bf{II}}}$ of $\bm{S}$ contains the complex amplitude reflection coefficients that describe internal reflection at the surface barrier from inside the crystal surface. This is found here for a one-dimensional barrier potential perpendicular to the surface of arbitrary shape by numerical solution of the wave equation.

In the thin layer of constant potential between the top atomic layer and the surface barrier are forward and backward travelling plane waves as described above that scatter between the barrier and crystal. In the simplified case \cite{mcrae1} of a single plane wave scattering in a crystal surface-projected band gap, the modulus of the crystal and barrier reflection coefficients is 1 indicating total reflection. Multiple scattering between crystal and barrier can occur and a standing wave can form for certain values of $(E, \bm{k}_{\parallel})$. This corresponds to the electron being confined in this region in a surface state. The condition for the formation of a standing wave is 
\begin{equation} \label{eq;phib}
\phi_b + \phi_c = 2 \pi n \;\;\;\;\; \text{with } n \text{ integer}
\end{equation}
where $ \phi_b$ and $\phi_c$ are the phase change on scattering at the barrier and crystal respectively. In terms of complex amplitude reflection coefficients of barrier $\rho_b$, and crystal $\rho_c$, this condition is
\begin{equation} \label{eq;rhob}
1 - \rho_b \rho_c = 0.
\end{equation}
where $\rho_b=\exp(i\phi_b)$ and $\rho_c=\exp(i\phi_c)$. For $(E,\bm{k}_{\parallel})$ near surface-projected bulk-gap edges $\phi_c$ varies rapidly and in the barrier image tail $\phi_b$ changes rapidly allowing Eqs.~(\ref{eq;phib}) and (\ref{eq;rhob}) to be satisfied often in these cases. Surface resonances may occur when the modulus of the crystal reflection coefficient is less than 1 giving weaker reflection, and scattering does not occur in a surface-projected band gap. In this case Eq.~(\ref{eq;rhob}) becomes
\begin{equation} \label{eq;rhoarrow}
|1 - \rho_b \rho_c| \rightarrow \text{  minimum}.
\end{equation}

In the case of real crystals more than one plane wave must be considered and a matrix representation is required where the reflection coefficients of barrier and crystal are contained in the matrices $\bm{S}^{\text{\bf{II}}}$ and $\bm{M}$ respectively. McRae has determined \cite{mcrae1} that in this case one or other of the eigenvalues of the matrix $\bm{S}^{\text{\bf{II}}} \bm{M} $ must approach unity for the occurrence of a surface state or resonance. This can be expressed as
\begin{equation} \label{eq;lambda}
|1 - \lambda_v| \rightarrow \text{  minimum}.
\end{equation}
where $ \lambda_v$ is the $v$th eigenvalue of the matrix $\bm{S}^{\text{\bf{II}}}\bm{M}$. For computational purposes it is convenient to re-express this condition in the form 
\begin{equation} \label{eq;minimum}
|\text{det} \; [\bm{I} - \bm{S}^{\text{\bf{II}}}\bm{M}] | \rightarrow \text{  minimum}	                
\end{equation}
where $\bm{I}$ is the identity matrix. This is the condition for sustained multiple scattering between the crystal atomic layers and the surface barrier and corresponds to the electron being trapped in the surface region. Hence mapping the values of $(E,\bm{k}_{\parallel})$ for which Eq.~(\ref{eq;minimum}) applies gives the surface band structure of the crystal surface. 

Echenique, Pendry and co-workers \cite{echenique} have used Eq.~(\ref{eq;phib}) to explain the occurrence of surface image states that arise from the image tail of the surface barrier. Smith and co-workers \cite{smith} and Borstel and Th\"{o}rner and co-workers \cite{borstel} have also used this condition to calculate the energy of Shockley and image SSs in \textit{sp}-band gaps of a number of metals by calculating $\phi_c$ from parameterised nearly-free electron bands. This method is not applicable generally and because of the narrow energy range considered in this work, it can be expected to yield mostly trends in the properties of the surface barrier between different surfaces of the metals.

The eigenvalues of the reduced transfer matrix $\bm{R}$ for the bulk crystal give the band structure of the bulk crystal in the form $E(k_{\perp})$ for each value of $\bm{k}_{\parallel}$ where $k_{\perp}$ is complex and is the component of the crystal momentum $\bm{k}$ perpendicular to a 2D atomic layer that is a plane of the crystal structure. This corresponds to the complex bulk band structure in the form $E(\bm{k})  = E(\bm{k}_{\parallel}, k_{\perp})$ where the usual bulk Brillouin zone (BBZ) for values of $\bm{k}$ is replaced by a zone in the form of a right prism. This prism has polygon basal planes that each are the 2D Brillouin zone for $\bm{k}_{\parallel}$ values in a particular crystal plane and sides that extend from the central plane for $k_{\perp}$ values of $-\pi/d$ and $\pi/d$ where $d$ is the spacing between atomic layers parallel to a particular 2D crystal plane. Bulk energy bands have finite values for Re $k_{\perp}$  with Im $k_{\perp}  = 0$ and have propagating wave functions. As first described by Heine \cite{heine}, gaps between energy bands are spanned by ``real lines" that have ``real energy lines" for Re $k_{\perp}$ and loops/half-loops for Im $k_{\perp}$ and have wave functions that are attenuating in a direction perpendicular to the particular crystal plane since they have finite values for both Re $k_{\perp}$  and Im $k_{\perp}$. Band minima that have no lower energy maxima to which to join have real lines that extend to infinity. If the crystal is terminated so that the particular crystal plane becomes the surface of the semi-infinite crystal the bulk wave functions with finite Im $k_{\perp}$ that attenuate towards the interior of the crystal \textit{may} form the bulk part of a surface state or resonance. This will occur if the wave function amplitude and derivative from the bulk crystal can match the form of the wave function through the surface and barrier potential transition regions and into the vacuum side of the semi-crystal. In the present scattering method this is achieved by detecting sustained multiple scattering in the surface region via the condition in Eq.~(\ref{eq;minimum}). It is necessary to have an accurate form for the shape and height of the surface barrier potential including the position of the image tail and its saturation closer to the surface as well as realistic ion-core and interstitial potentials in the surface and bulk regions.

The surface energy band structure $E(\bm{k}_{\parallel})$ contains SS/SRs and the continuum of surface-projected bulk states from the $E(\bm{k}_{\parallel},k_{\perp})$ band structure. In the surface projection of bulk bands $E(k_{\perp})$ for values of $\bm{k}_{\parallel}$, absolute, symmetry or partial bulk energy-gaps arise from the critical points (maxima and minima) of the bulk bands. Surface-projected partial bulk energy-gaps occur where projected bulk band gaps are also overlaid with propagating bulk bands. Surface states lie in energy ranges of symmetry or absolute surface-projected bulk-band gaps below the vacuum level, $E_v$. Surface resonances lie in ranges of surface-projected partial (quasi) bulk gaps below $E_v$ or they have energies above $E_v$ where they are degenerate with vacuum states. One very useful expression for the wave function of an SS/SR in terms of an expansion in terms of bulk states that are represented in the linear-combination-of-atomic-orbitals approximation, is given by Bertel \cite{bertel} as 
\begin{widetext}
\begin{equation} 
\Psi_{\bm{k}_{\parallel}}(\bm{r}) =  \sum_j e^{i\bm{k}_{\parallel}.\bm{R}_j}  \sum_{k_{\perp}}  e^{i \kappa.\bm{R}_j}  \sum_n e^{\mu z} a_{n,k_{\perp}}  \sum_l b_{nl} \chi_l (\bm{r} - \bm{R}_j).
\end{equation}
\end{widetext}

The factor $\sum_l b_{nl} \chi_l (\bm{r} - \bm{R}_j)$ gives the atomic-orbital composition (\textit{s}, \textit{p} etc) centred at lattice sites $\bm{R}_j$ of bulk band $n$. The $+z$ coordinate is the inward normal to the surface and $k_{\perp}  = \kappa + i \mu$. The third factor gives the relative contribution to the wave function from all $k_{\perp}$ values from all bulk bands $n$ together with the exponential damping factor of that contribution, Im $k_{\perp}  = \mu$ towards the interior of the crystal from the real lines and $\mu$ must be negative in the present scheme. A small (large) value of $|\mu|$ indicates a long (short) decay length for each contribution to the wave function and each contribution decays by a factor $\exp(- |\mu| d)$ between atom surface layers with spacing $d$ normal to the surface. The contribution that is the most slowly attenuating in space has the smallest value of $|\mu|$, $\mu_{\text{min}}$ and $\lambda_s = 1/\mu_{\text{min}}$ is often taken as a measure of the decay length of the SS/SR in the bulk.

The complex electron self-energy $\Sigma$ expresses the many-body dynamic electron-electron interaction. For calculating bulk and surface excited states with higher electron energies, the $E$ and $\bm{k}$ variation of $\Sigma$ from ground state values becomes significant and must be taken into account. \cite{jepsen} Re $\Sigma$ alters the energy position of one-electron bands and Im $\Sigma$ modifies the bands due to inelastic processes from single particle and surface and bulk plasmon excitations. These variations can be included in a phenomenological way. Inelastic effects due to surface and bulk electron-phonon interaction can also be included via imaginary contributions to the scattering phase shifts. \cite{jepsen}

The scattering solution for the electron with $(E, \bm{k}_{\parallel})$ has the wave function of the electron in the thin 2D periodic layer of constant potential between the top atomic layer and the surface barrier described by forward and backward travelling plane waves that scatter between the barrier and crystal atomic layers. For the case of negligible inelastic electron-electron interaction, the respective components of their wave vectors, $\bm{k}_v^{\pm}$ parallel to the surface and along the inward surface normal of the crystal, for each plane wave $v$ are given by \cite{mcrae2}
\begin{eqnarray} \label{kperp}
\bm{k}_v^{\pm} & = & (\bm{k}_{\parallel}  + \bm{v}, \pm k_{\perp,\bm{v}})  \text{  and} \\	
k_{\perp,\bm{v}} & = & (E  - | \bm{k}_{\parallel}  + \bm{v} |^2)^{1/2}      \text{ (positive root,} \\
& & \text{\hspace{2.95 cm} real or imaginary)} \notag	
\end{eqnarray}
using Rydberg atomic units (1 Ry = 13.6 eV, 1 a.u. =  0.5292~\AA). Plane waves directed inward towards the crystal have $\bm{k}_v^+$ and $\bm{v}$ labels the reciprocal-net vectors of the surface. In the present case we have no variation of the potentials of surface layers of atoms with respect to bulk layers. The surface barrier is one-dimensional with only variation in the direction normal to the surface. The height of the barrier, $U_0$ is $E_v-E_c$. The determination of the complex reflection coefficients for scattering at this surface barrier involves only the energy associated with the motion perpendicular to the surface $E_{\perp,\bm{v}} = (\text{Re } k_{\perp,\bm{v}})^2$ and the wave vector normal components $\pm k_{\perp,\bm{v}}$ of the plane waves. 

Historically surface states (and resonances) have been designated as Shockley type or Tamm type. Shockley type occur when bulk bands cross on forming surface-projected band gaps and the wave functions interact causing a hybridized charge density. They are sensitive to the surface barrier potential shape and do not require an unsymmetrical termination of the bulk periodic part of the potential at the surface. Tamm type can occur when there is an unsymmetrical termination of that periodic potential at the surface and do not require band crossing. They retain the same characteristics as the band from which they derived.

In a scattering picture for this present model there are only two differentiated ways of forming an SS/SR. If the sustained multiple scattering condition of Eq.~(\ref{eq;minimum}) involves some of the plane waves of the set with finite Re $k_{\perp,\bm{v}}$ and $E_{\perp,\bm{v}}$ reflecting from points along the shape of the surface barrier then the SS/SR main wave function amplitude may extend well past the centre of the top atomic layer of atoms into the vacuum. These SS/SRs are sensitive to the surface barrier shape. Also included here are Rydberg or image SS/SRs where reflection occurs at the image tail of the surface barrier. Here the SS/SR main wave function amplitude extends even further into the vacuum. We term these as surface barrier or type-one SS/SRs. The special case of the Shockley SS/SRs is of this type.

If the condition of Eq.~(\ref{eq;minimum}) involves only those plane waves of the set with Re $k_{\perp,\bm{v}}  = 0 = E_{\perp,\bm{v}}$ (only finite Im $k_{\perp,\bm{v}}$) reflecting from the bottom of the barrier then the SS/SR main wave function amplitude does not extend far beyond the top row of atoms into the vacuum. These particular plane waves are exponentially attenuating or evanescent. The bottom of the barrier corresponds in energy to the constant potential energy of the top atomic layer and spatially is where this potential just starts to turn upward to produce the barrier potential. Here the SS/SRs arise because of the unsymmetrical termination of the bulk periodic part of the potential at the surface due to the overlapping up-turn of the potential of the surface barrier. These SS/SRs are not sensitive to the surface barrier shape but only the start of the overlapping potential up-turn. We term these as type-two SS/SRs and they are more Tamm-like.

For the scattering method here it is apparent that labelling an SS/SR as either Shockley or Tamm type is only possible in some special cases. 

The calculation of the matrices $\bm{R}$, $\bm{M}_1$, $\bm{M}$ and $\bm{S}^{\text{\bf{II}}}$ is the same as that used by McRae in the calculation of intensities in low-energy electron diffraction (LEED). \cite{mcrae2}

\section{Recent surface band calculations for Al(111)}

The most complete and recent calculations of surface band structure for Al(111) are an \textit{ab initio} self-consistent finite-layer slab calculation using the density functional (DF) method with local-density-approximation (LDA) \cite{heinrich1} that was also extended to higher energies using a non-local exchange-correlation (xc) potential to produce the image tail of the surface barrier potential \cite{heinrich2}; an \textit{ab initio} self-consistent surface-embedded Green-function (SEGF) method for a semi-infinite crystal \cite{benesh} and an \textit{ab initio} self-consistent assembly of boundary-controlled monolayers (ABCM) method \cite{hummel} for a semi-infinite crystal.

The DF calculations allowed for free-form surface potentials, the SEGF method used a full-potential with warping terms in the interstitial region and non-spherical terms inside the atomic muffin-tins and the ABCM method used the warped-muffin-tin approximation where there is no restriction on the shape of the self-consistent potential outside the muffin-tin spheres.

From Fig.~5 of Ref.~\onlinecite{hummel} we note that the valence charge density of the sub-surface atomic layers is the same as the bulk, and as noted in Ref.~\onlinecite{hummel} ``considerable charge transfer is essentially limited to the vacuum side of the ion cores and the charge density on the bulk side is little different from the bulk". The charge density shown in Fig.~2 of Ref.~\onlinecite{benesh} is similar to that of Ref.~\onlinecite{hummel}.

\section{Determination of bulk, surface and barrier potentials.}

The lattice constant for fcc Al at room temperature of 300 K is 4.0496~\AA\ and the (111) surface has primitive net vectors of length 2.863~\AA\ at an angle of $120^{\circ}$. The spacing between bulk layers parallel to the surface is $d = 2.338$~\AA. The primitive reciprocal-net unit cell is rhombic with sides of length 2.534~\AA$^{-1}$ at $60^{\circ}$ and the surface Brillouin zone (SBZ) is hexagonal. Recent LEED determinations of surface structure have found an expansion of the surface layer position normal to the surface of $\sim 1.4$ \% \cite{burch} and smaller variations from bulk values for sub-surface layers. These variations are considered to have a negligible effect in the present calculation and for clarity in the analysis of results they are not included here. For the same reasons electron-phonon interactions at 300 K are also not included. 

The ion-core scattering phase shifts were calculated from the bulk fcc Al potential found from the self-consistent band structure calculation of Moruzzi et al.\cite{moruzzi} This potential is of the muffin-tin form with the self-energy xc potential calculated from the Hedin-Lundqvist LDA functional. The muffin-tin radius is $1.42$~\AA. The Fermi energy, $E_f$ is 8.6 eV above the bulk constant potential energy, $E_c$. The symmetry designations of all the bands are shown in the bulk band structure calculation of Ref.~\onlinecite{szm}. More recent \textit{ab initio} calculations have been performed for the bulk band structure of Al with full potentials that include non-muffin-tin corrections and methods to account for the self-energy $\Sigma$ from many-body electron-electron interaction. The self-consistent DF calculation using the CRYSTAL software with use of a local and non-local xc functional gave results extending to 14 eV above $E_f$. \cite{jacobs} A DF calculation using the ABINIT software incorporated a many-body perturbation approximation to account for the self-energy. This GW approximation used the Green function propagator G and static self-consistent screening interaction W to give results to 12 eV above $E_f$. \cite{bruneval} For energies up to 9 eV above $E_f$ both of the latter calculations show small differences from each other and small differences with the band structure from Ref.~\onlinecite{moruzzi} and that calculated in the present work except for the \textit{sd}-like $X_1^{\ell}$ point. The above two DF calculations place the $X_1^{\ell}$ point  $\sim 0.5$ eV higher than the other two calculations.

The surface and bulk plasmon excitation energies for this surface are at $10.2 \pm 0.2$ eV and $15.0 \pm 0.2$ eV \cite{levinson} above $E_f$ where inelastic electron-electron interaction becomes significant. Below these energies the effects of inelastic processes on bands from Im $\Sigma (E, \bm{k})$ are negligible and are not included in the present calculation except as a check in one particular case as noted later. Likewise the corrections necessary for excited high-energy bands from Re $\Sigma (E, \bm{k})$ are not included here because the present energy range is only 9 eV above $E_f$.

For simple metals \cite{riffe,alden} the reduced atom coordination at the surface results in band narrowing in the surface layer, a decrease in surface ion-core binding energies and decrease in the magnitude of the average potential in the surface layer. The spill-out of electrons at the surface results in the opposite change in the ion-core binding energies and magnitude of the average potential in the surface layer. \cite{riffe,alden} The experimental surface core-level (binding-energy) shifts (SCLS) measure the initial-state contribution from the net above effects and also a final-state contribution from the relaxation energy due to the change in screening between bulk and surface. For simple metals the final-state contribution is small \cite{riffe,alden} and hence the experimental SCLS are approximately equal to the initial state SCLS. Ald\'{e}n et al \cite{alden} have found that the initial state SCLS for simple metals is approximately equal to the change in magnitude of the surface-layer average potential energy. For Al(111) the experimental SCLS of the 2p core state for the first and second surface atom layers are only $- 0.027 \pm 0.03$ eV and 0 respectively. \cite{borg} Hence assuming the final-state SCLS to be small we approximate the surface ion-core potential to be the same as the bulk and also the interstitial potential in the surface layer to be the same as the bulk up to the centre of the top row of atoms. Past this point we add the one-dimensional surface barrier potential that arises from the spill-out of the valence electrons into the vacuum region. It is to be noted that this effective overlapping of the spherically symmetric muffin-tin potential at the surface and the surface barrier normal to the surface gives a non-spherical contribution to the surface potential. This model is also consistent with the charge density results from self-consistent calculations for Al(111) using free-form surface potentials that were discussed above in Sec.~III. 

The experimental work function $\Phi$ for Al(111) is $4.24 \pm 0.02$ eV \cite{grepstad} and this gives the vacuum level $E_v$ with respect to $E_c$ as $E_v - E_c = E_f + \Phi = 12.84$ eV. We use an empirical surface barrier shape $U_b(z)$ that has been suggested by Rundgren and Malmstr\"{o}m \cite{rundgren} where $z$ is the perpendicular distance from the centre of the first row of atoms which is located at $z = 0$ and the barrier is located at negative $z$ values. In Rydberg atomic units it is
\begin{widetext}
\begin{equation} \label{eq;Ub}
U_b(z) =  
\left\{
\begin{array}{lcl} 
E_v + 1/[2(z - z_0)] & \text{  for  } & z < z_1 < z_0, \\
v_0 + v_1(z - z_1) + v_2(z - z_1)^2 +v_3(z - z_1)^3 &  \text{  for  } & z_1 < z < z_2, \\
 E_{c,1} & \text{  for  } & z > z_2.	
\end{array}
\right.
\end{equation}
\end{widetext}
It has an image tail of the form $1/[2(z-z_0)]$ where $z_0$ is the position of the image plane. This joins smoothly to a cubic-polynomial-type saturation at a point $z_1$ closer to the metal surface and also smoothly to the constant potential of the surface layer, $E_{c,1}$ at $z_2$. Here we have set the barrier to join to the constant potential energy $E_{c,1}$ at the centre of the atoms in the top layer at $z_2$ = 0. With the same constant potential for the surface layer as the bulk and energies with respect to $E_c$, the height of the surface barrier $U_0$ = $E_v$ - $E_c$ = 12.84 eV. From the smooth joining of the segments and that $U_0$ here is pre-determined from the bulk band structure and experimental work function we have only two independent parameters - the image plane $z_0$ and the point of onset of saturation $z_1$. The parameters $z_0$ and $z_1$ are to be determined from the experimental energies of SS/SRs that depend on the shape of the surface barrier potential at $\overline{\Gamma}$.

At $\overline{\Gamma}$ an SS has been detected by high-resolution ARPES at $4.56 \pm 0.04$ eV below $E_f$. \cite{kevan} The first Rydberg image resonance has been detected at 3.75 eV above $E_f$ or $\sim 0.5$ eV below the vacuum level by k-resolved inverse photoemission spectroscopy (KRIPES) \cite{heskett,yang} and scanning tunnelling spectroscopy (STS) \cite{yang} and by two-photon photoemission (2PPE) \cite{bulovic} at $0.46 \pm 0.1$ eV below the vacuum level.

The shape details of the empirical surface barrier that are to be used in the calculations for the whole SBZ are found by varying the parameters $z_0$ and $z_1$ to obtain a match between the calculated and experimental surface bands at $\overline{\Gamma}$.

\section{Calculation of surface states/resonances for $\bm{\overline{\Gamma}}$.}

\begin{figure}
\includegraphics[scale=0.475]{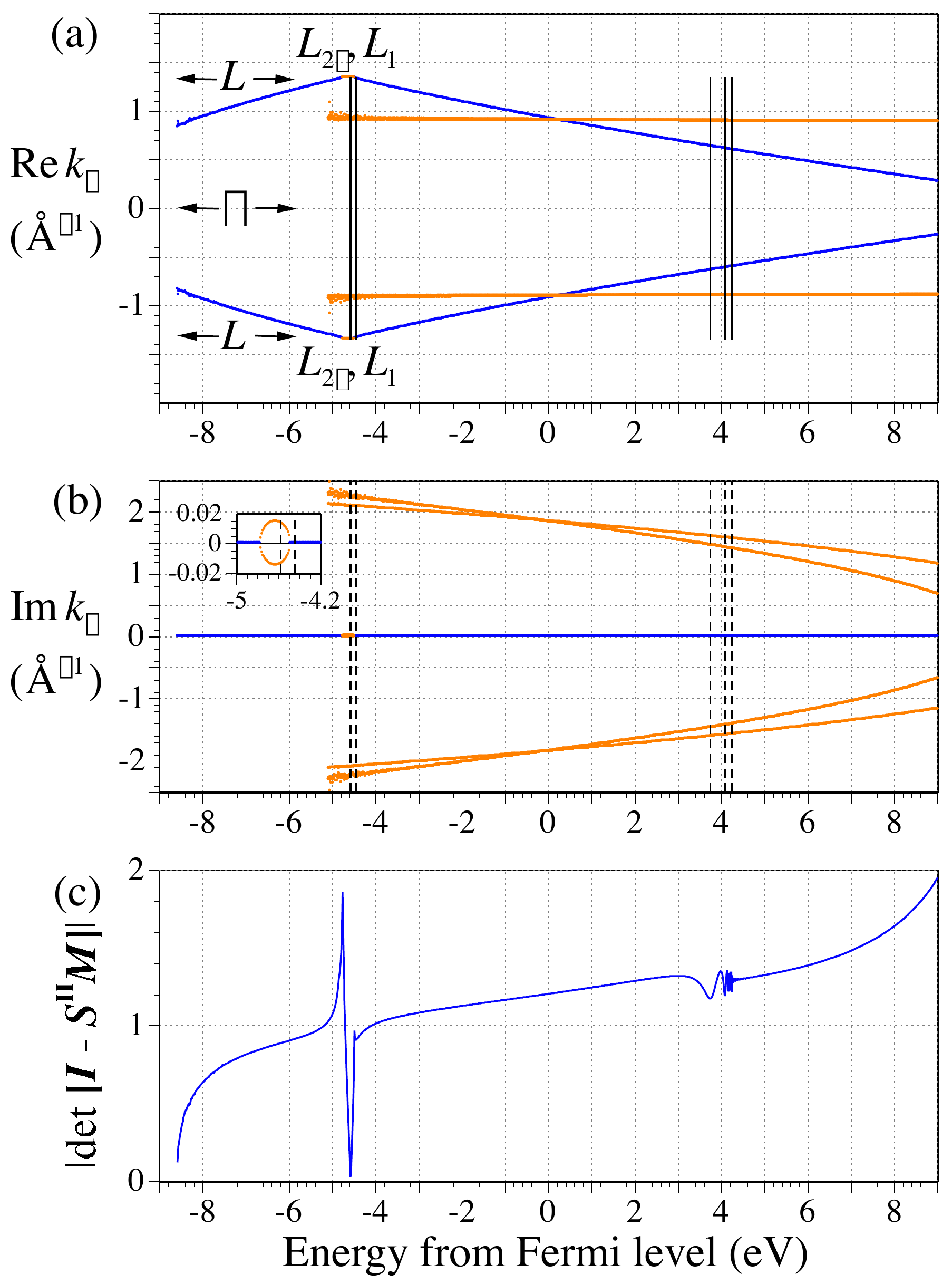}
\caption{(color online) The real and imaginary parts of the complex bulk and surface energy band structure $E(\bm{k}_{\parallel},k_{\perp})$ at $\bm{k}_{\parallel} = 0 \equiv \overline{\Gamma}$ for Al(111) are shown in panels (a) and (b) respectively. These $\bm{k}$ values pass along a line through the $\Gamma$ and $L$ points of the BBZ as indicated in panel (a). Propagating bulk bands are blue/grey points and real lines of attenuating bulk bands are orange/light grey points. The real lines that extend to $-\infty$ from `unmatched' bulk-band minima are terminated in the diagram at low energies where they become unstable and have negligible effect in the calculation. Other real lines from such minima at higher energies that do not produce a significant effect in the calculation and have large Im $k_{\perp}$ values are not shown. Bulk band labels at $L$ are indicated at the margins. The insert in panel (b) shows an expanded scale for the energy range $-5$ to $-4.2$ eV. Note that loops/half-loops in panel (b) show most readily the surface-projected absolute and partial bulk band gaps. Vertical full black lines indicate energies of surface states or resonances. The intersection of vertical dashed black lines in panel (b) with bulk Im $k_{\perp}$ values indicates attenuation factors $|\mu|$ of each bulk-band contribution to the surface state or resonance. Near 4 eV are the first, second and last ($n=\infty$) members of the Rydberg series of image surface resonances. Panel (c) shows the result of Eq.~(5) for the above $\bm{k}_{\parallel}$ value where minima indicate energies of surface states and resonances. The minium at $-8.6$ eV corresponds to a surface resonance for a free electron in the crystal.}
\end{figure}

First we calculate the complex bulk band structure in the form $E(\bm{k}) = E (\bm{k}_{\parallel}, k_{\perp})$ where $\bm{k}_{\parallel} = 0$ corresponds to $\overline{\Gamma}$ of the SBZ from the bulk potential from Ref.~\onlinecite{moruzzi}. This complex bulk band structure also shows attenuating bulk states. The energy interval in all the calculations in this work is 0.01 eV. Seven plane waves were found to be adequate for the plane-wave basis set for interlayer scattering at $\overline{\Gamma}$. Only the plane wave with reciprocal-net-vector coefficients 00 is propagating in this energy range and the 3-fold degenerate $\{0\overline{1}\}$ and $\{\overline{1}0\}$ sets are evanescent plane waves. Spherical partial waves up to $l_{\text{ max}} = 5$ were sufficient to describe the intralayer scattering. The resulting band structure is shown in Fig.~1(a) and (b) and passes through the $\Gamma$ and $L$ points of the BBZ as shown. All energies are subsequently referred to $E_f$ unless stated otherwise. Hence we see that the gap between bulk bands (blue/grey) at $L_{2'}$ and $L_1$ near $- 4.6$ eV projects as an absolute surface gap at $\overline{\Gamma}$. We also note the real lines with real energy lines (orange/light grey) in Fig.~1(a) in the $\textit{sp}$ $L$ gap and small loop (orange/light grey) in Fig.~1(b) that is also shown in the insert on an expanded scale. Bands (blue/grey) in Fig.~1(a) give lines at Im $k_{\perp} =0$ only (blue/grey) in Fig.~1(b). We also note the real energy lines and half-loops that start from bulk minima at higher energies and extend to $-\infty$. These are from minima in $\Lambda_3$ and $\Lambda_1$ bands that are shown in Ref.~\onlinecite{szm}. These real energy lines and half-loops indicate a very wide projected partial even-symmetry gap from the $\Lambda_1$ minimum ($15.5$ eV) to $-\infty$ and a wide odd-symmetry gap from the $\Lambda_3$ minimum ($10.5$ eV) to $-\infty$. Other real energy lines and half-loops with larger Im $k_{\perp}$ values for the present energy range from higher energy minima are not calculated because they have negligible effect at these lower energies and also lead to instabilities in the calculation. 

Using Eq.~(\ref{eq;minimum}) we calculate the energies of the SS/SRs at $\overline{\Gamma}$ where $\bm{k}_{\parallel} = 0$ and find that the first Rydberg image SR energy is reproduced at $- 0.50$ eV with respect to $E_v$ with $z_0$ at $-0.53$~\AA\ and $z_1$ at $-1.06$~\AA. Also without any change in these $z_0$ and $z_1$ values the lower SS is at $-4.58$ eV and in the correct experimental energy range. At the condition of Eq.~(\ref{eq;minimum}), only the 00 plane wave is propagating with finite Re $k_{\perp}$ and is incident at points along the shape of the surface barrier and reflects from it. Hence these SS/SRs are type one. The SS conforms to the classical Shockley type. It is placed slightly closer to the $L_1$ gap edge in the $L_{2'}$, $L_1$ absolute gap with width calculated here to be 0.29 eV. The position of the SS with respect to gap edges is not certain experimentally and there are no experimental determinations of the energy positions of the $L_{2'}$, $L_1$ points. \cite{kevan} As the accuracy of the calculated positions of the $L_{2'}$, $L_1$ points from the present band structure cannot be assessed we have not attempted to also position the SS with respect to either gap edge. The result from Eq.~(\ref{eq;minimum}) is shown in Fig.~1(c). This result was found previously where it was used to determine SRs above the vacuum level to 27 eV at $\overline{\Gamma}$ where inelastic scattering from Im $\Sigma$ is significant and must be included. \cite{read80} Here we are interested in extending the calculation to all $\bm{k}_{\parallel}$ points of the SBZ but for energies where Im $\Sigma$ is negligible. Scattering phase shifts at the image part of the barrier in multiples of $2 \pi$ from that found for the above $z_0$ will position the image SRs at almost the same energy. The values $z_0 = -2.70$~\AA\ and $z_1 = -4.2$~\AA\ also produce the correct first Rydberg image SR energy and also the correct SS energy. However these values also produce an SR at 0.60 eV at $\overline{\Gamma}$ for this wider barrier. At present there is no experimental evidence for such an SR. Hence the first values for the shape of the surface barrier were used in the present calculation. This barrier shape is plotted in Fig.~2.

\begin{figure}
\includegraphics[scale=0.465]{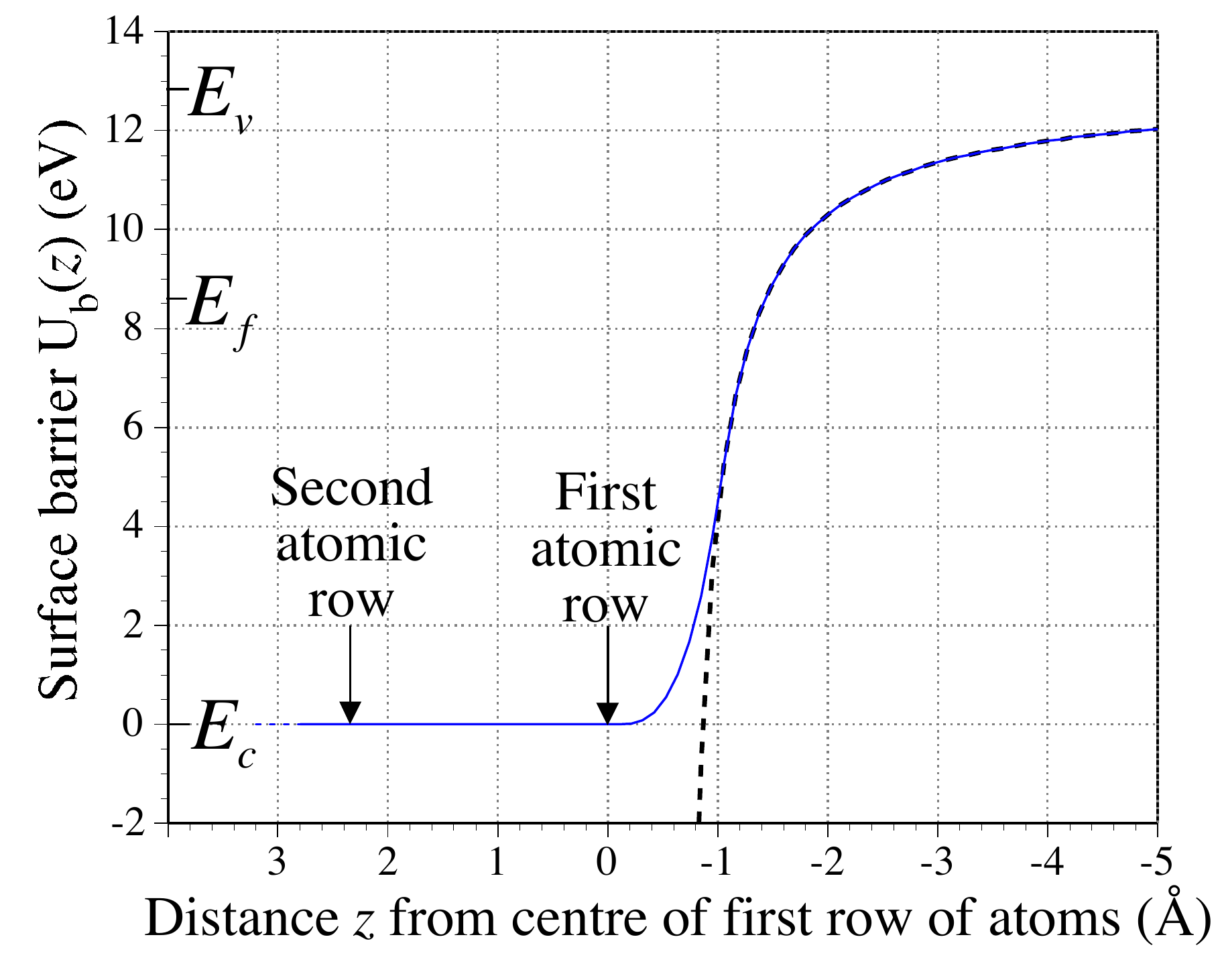}
\caption{(color online) Surface barrier potential energy relative to $E_c$, the bulk constant potential energy for Al(111). The dashed line is the continuation of the unsaturated image potential. The image plane position $z_0 = -0.53$~\AA\ and saturation begins at $z_1 = - 1.06$~\AA. The muffin-tin radius is 1.42 \AA\ and the geometric surface $z_j = - 1.17$~\AA. The vacuum and Fermi level are $E_v$ and $E_f$ respectively relative to $E_c$.}
\end{figure}

\begin{figure*}
\includegraphics[scale=1.35]{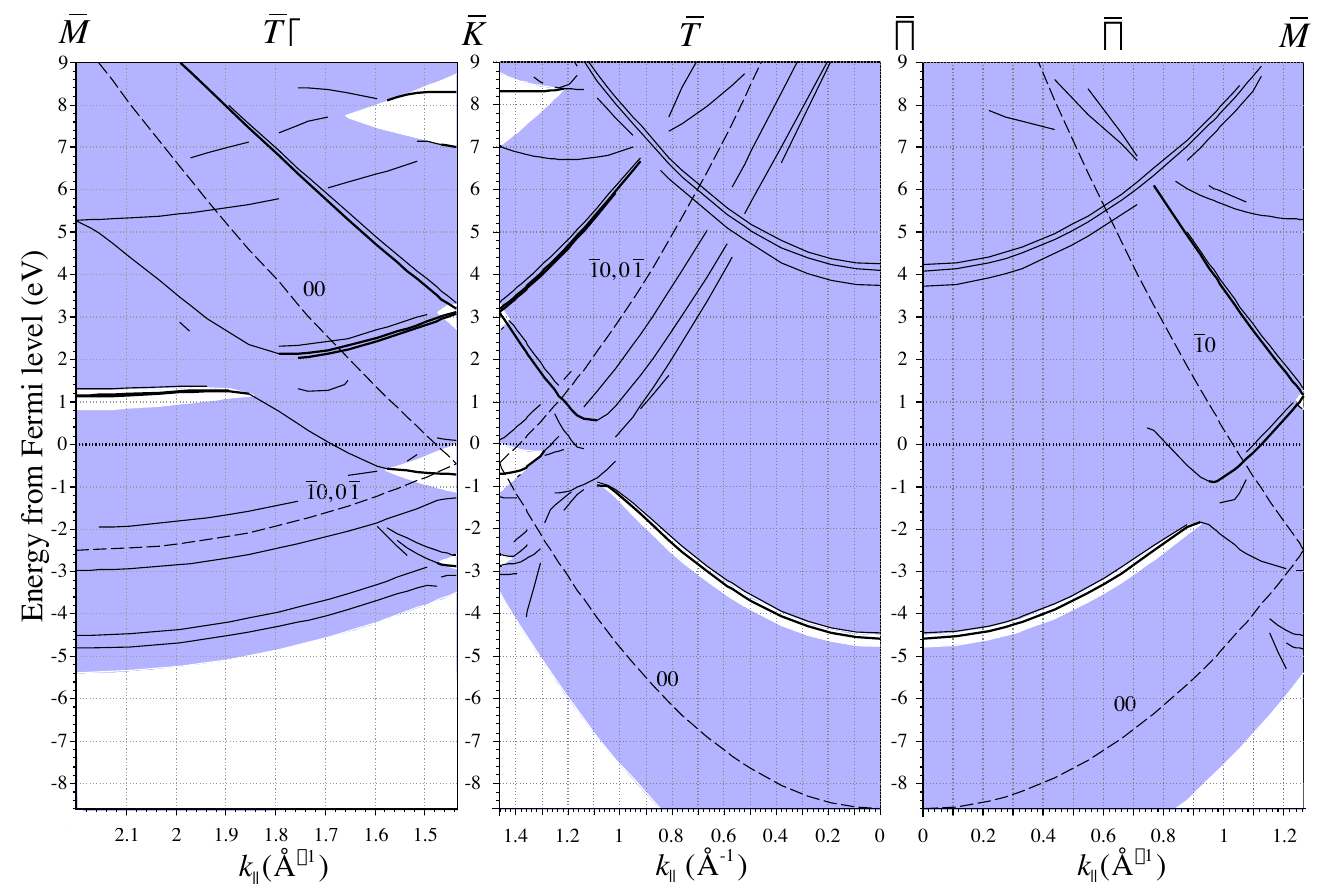}
\caption{(color online) Surface energy band structure $E(\bm{k}_{\parallel})$ for Al(111). Surface states and resonances are indicated by full black lines. Thick full lines indicate surface states or (sections of) strong surface resonances where substrate reflection is strong. Thinner full lines indicate (sections of) weaker surface resonances. The continuum of surface-projected bulk bands are shaded blue/grey and their absolute gaps unshaded. Dashed black lines indicate the free-electron surface band structure labeled with the coefficients of the surface reciprocal-net vectors. All SS/SRs below the lowest energy free-electron band are necessarily type two. The horizontal axes are scaled differently in order to display the relevant features more clearly.}
\end{figure*}

The above calculated energy positions of the SS and SR, the first, second and last ($n=\infty$) Rydberg image SRs and projected band gap for $\overline{\Gamma}$ are plotted on the surface band structure $E(\bm{k}_{\parallel})$ for $\overline{M}$ -- $\overline{K}$ -- $\overline{\Gamma}$ -- $\overline{M}$ shown in Fig.~3. The SS/SRs are also plotted in Fig.~1(a) (vertical black lines) where they have no dispersion with values of Re $k_{\perp}$, to show the full surface and bulk band structure at $\overline{\Gamma}$. These full band structures for a given value of $\bm{k}_{\parallel}$ show surface, propagating bulk and also attenuating bulk bands near the surface and all are important in the analysis of energy and momentum distribution curves in photoemission and in other surface spectroscopies. The attenuation constants $|\mu|=|\text{Im }k_{\perp}|$ of each bulk-band contribution to the SS/SRs are indicated from the SS/SR energy intersection (vertical dashed black line) with bulk Im $k_{\perp}$ values. The small values of Im $k_{\perp}$ in the loop in Fig.~1(b) for the $L$ gap and at the energy of the SS account for the fact that this SS has a wave function component with large decay length $\lambda_s$ and long penetration into the bulk of the crystal as well as components that do not penetrate much past the surface layer. From Fig.~1(b) $\lambda_s = 1/\mu_{\text{min}} \approx 71$~\AA\ in the present calculation. This SS is related to the SS found experimentally on Al(100) at $-4.55$ eV at $\overline{X}$ in the absolute $L$ gap. \cite{levinson,kevan} For the bulk and surface potentials used here the SS lies slightly closer to the upper $L_1$ (\textit{sd}-like \cite{bertel}) band edge at $-4.49$ eV than $L_{2'}$ ($p$-like \cite{bertel}) at $-4.78$ eV and there is an SR just above the $L_1$ gap edge at $-4.45$ eV. If the surface barrier parameters are changed slightly so that the SS is formed nearer the lower energy gap-edge then this extra SR changes position to just below the $L_{2'}$ edge. This is a very surface localised SR with calculated $\lambda_s \approx 0.5$~\AA\ that may affect photoemission energy and momentum distribution curves. This SR is in the very wide partial projected band-gap formed between the band minima from the higher energy $\Lambda_1$ bulk band and $- \infty$ that is mentioned above. The Rydberg image SRs are also in this wide partial gap with calculated $\lambda_s \approx 0.7$~\AA. This type of partial projected gap will be discussed again later. The separation between the surface and sub-surface atomic layers here is 2.34~\AA.

\section{Calculation of surface states/resonances at $\bm{\overline{M}}$.}

\begin{figure}
\includegraphics[scale=0.48]{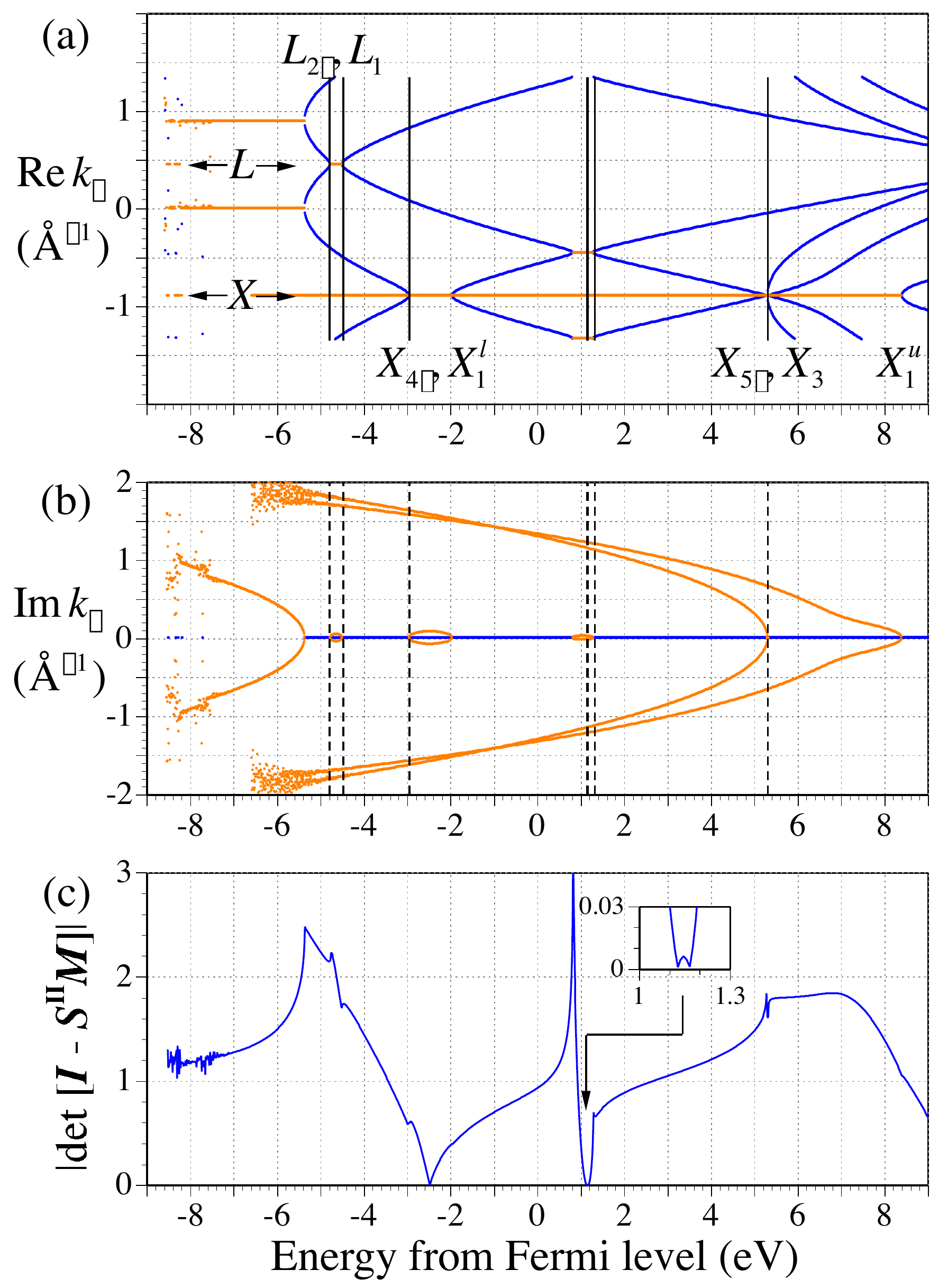}
\caption{(color online) The real and imaginary parts of the complex bulk and surface energy band structure $E(\bm{k}_{\parallel},k_{\perp})$ at $\bm{k}_{\parallel} \equiv \overline{M}$ for Al(111) are shown in panels (a) and (b) respectively. These $\bm{k}$ values pass along a line through the $L$ and $X$ points of the BBZ as indicated in panel (a). Bulk band labels at $L$ and $X$ are indicated at the margins. A very small gap at 5.3 eV between the band maximum at $X_{5'}$ and the band minimum at $X_3$ is not discernable on the scale of the diagram. Other features of panels (a) and (b) are the same as Fig.~1. Panel (c) shows the result of Eq.~(5) for the above $\bm{k}_{\parallel}$ value where minima indicate energies of surface states and resonances. The insert shows an expansion for the energy range 1 to 1.3 eV. The minium at $-2.49$ eV corresponds to a surface resonance for a free electron in the crystal. }
\end{figure}

The calculated complex bulk band structure for $E(\bm{k}_{\parallel}, k_{\perp})$ with $\bm{k}_{\parallel}= 1.27$ \AA$^{-1} \equiv \overline{M}/\overline{M}' (\overline{M}(L)/\overline{M}(X))$ point is shown in Fig.~4(a) and (b) that also shows attenuating bulk states. Note that loops/half-loops in panel (b) show most readily the surface-projected absolute and partial bulk band gaps. We find convergence with up to eight plane waves in our region of constant potential between scattering layers. Here the bulk bands pass through the $L$ and $X$ points of the BBZ as shown. The $L_{2'}$, $L_1$  bulk gap near $-4.6$ eV becomes a surface-projected partial bulk band gap at $\overline{M}$ because it is overlaid with other free-electron-like bulk bands in the surface projection. The bulk gap between the $X_{4'}$ point at $-2.98$ eV and $X_1^{\ell}$ point at $-1.98$ eV is also a projected partial gap overlaid with free-electron-like bulk bands. Real lines spanning these bulk gaps are real energy lines in Fig.~4(a) and loops in Fig.~4(b). Energy bands in Fig.~4(a) give lines at Im $k_{\perp} = 0$ only in Fig.~4(b) that also pass through loops in the case of projected partial gaps. The 2-fold degenerate $X_{5'}$ point is at 5.30 eV and the $X_3$ point is slightly higher with a very small $0.02$ eV gap between these points giving a projected partial gap between 5.30 and 5.32 eV. The real energy line and loop for this partial gap are not discernable on the scale of these diagrams. The $X_1^u$ point is at 8.40 eV. From $X_{5'}$ in Fig.~4(a) the band that has a maximum there and disperses downward has even bulk symmetry and the other flatter band that is a minimum there and disperses upwards has odd bulk symmetry. The band minima at $X_1^u$ (8.40 eV) and at $X_{5'}$ (5.30 eV) from the flatter band have no maxima to which to join and they produce real energy lines for Re $k_{\perp}$ and half-loops for Im $k_{\perp}$ extending to $- \infty$. Since all bands below $X_{5'}$ have even bulk symmetry, a wide projected odd-symmetry bulk gap exists from $X_{5'}$ to $- \infty$ and a wide even-symmetry partial bulk gap from  $X_1^u$  to $- \infty$. An absolute projected gap exists near 1 eV and below the band minimum at $-5.37$ eV.

The result of the calculation for the SS/SRs from Eq.~(\ref{eq;minimum}) is shown in Fig.~4(c) and energies plotted in Fig.~4(a) \& (b) and in Fig.~3 for $\overline{M}$ together with the absolute projected gaps. The minimum at $- 2.49$ eV in Fig.~4(c), as well as the minimum at $-8.6$ eV at $\overline{\Gamma}$ in Fig.~1(c), are not SRs of this system but indicate the crystal empty-net or surface free-electron energy. This occurs when a plane wave in the region of constant potential between layers transitions from evanescent to propagating. These energies are also plotted in Fig.~3. Above $-2.49$ eV only the two degenerate plane waves with reciprocal-net-vector coefficients 00 and $\overline{1}0$ become propagating in this energy range. Below $-2.49$ eV these plane waves and all others are evanescent and have Re $k_{\perp} = 0$ and finite Im $k_{\perp}$. The results in Fig.~4(c) determine that there are three occupied SRs below $-2.49$ eV at $\overline{M}$. Two are at $-4.80$ and $-4.48$ eV which are near the energies of the partial gap near $-4.6$ eV with edges at $L_{2'}$ ($-4.78$ eV) and $L_1$ ($-4.49$ eV). Both of these SRs are just outside the edges of this partial gap and have small $\lambda_s$ values $\approx 0.6$~\AA. However a small change in the partial gap position and width from the bulk band structure could put one or both of these SRs inside the partial gap and consequently with a large $\lambda_s$ value. Another SR is at $-2.96$ eV in the partial gap with edges $X_{4'}$ at $-2.98$ eV and $X_1^{\ell}$ at $-1.98$ and 0.02 eV above the lower energy p-like $X_{4'}$ edge. The calculated $\lambda_s$ is $\approx 53$~\AA. This SR is related to the SS observed on Al(100) at $-2.75$ eV at $\overline{\Gamma}$ in the absolute $X$ gap and $\sim 0.1$ eV above the $X_{4'}$ energy. \cite{hansson,levinson,kevan} The three SRs here arise from scattering at the bottom of the surface barrier and are not sensitive to its shape but are slightly sensitive to its starting position $z_2$. According to our classification in Sec.~II they are type-two SRs with wave function maximum amplitude not extending far beyond the top row of atoms into the vacuum.

There is no identification of occupied SRs at $\overline{M}/\overline{M}'$ in the ARPES experimental data of Hofmann and Kambe \cite{hofman} and no other experimental data at $\overline{M}/\overline{M}'$.

Above $-2.49$ eV the $00$ and $\overline{1}0$ plane waves have finite Re $k_{\perp}$ and  are incident at points along the shape of the surface barrier potential and reflect from it. The absolute gap near 1 eV has edges at 0.80 and 1.26 eV and has two SSs at 1.14 eV ($\lambda_s \approx 41$~\AA) and 1.17 eV ($\lambda_s \approx 48$~\AA) and there is an extra SR just above this gap with small $\lambda_s$. This SR is in the wide partial gap between the band minimum at $X_1^u$ and $- \infty$. There is also an SR at 5.30 eV in the very small $X_{5'}$, $X_3$ partial gap with large $\lambda_s$.

A KRIPES experimental study of unoccupied surface bands in this region found an SR at 4.0 eV for $\overline{M}'$ \cite{yang2} whereas we calculate an SR at 5.30 eV. Hence there is a 1.3 eV discrepancy. $\overline{M}$ and $\overline{M}'$ are different only with respect to possible allowed initial and final state transitions in (inverse) photoemission. The bulk band structure in Figs.~4(a) and (b) for $\overline{M}$ differs from that of $\overline{M}'$ only that the sign of $k_{\perp}$ values are interchanged. The partial gap at $\overline{M}$ and $\overline{M}'$ near 5.3 eV that we calculate is at the same energy as the partial gap in Fig.~6(b) at $\overline{M}'$ in Ref.~\onlinecite{yang2}. We also find the dispersion of this gap for $\overline{\Gamma}$ -- $\overline{M}'$ to be the same but Fig.~6(b) of Ref.~\onlinecite{yang2} shows a projected gap for $\overline{M}'$ at $\sim 1.8$ eV whereas ours is centred at 1.0 eV. The Al band calculations mentioned previously that use corrections for electron self-energy $\Sigma$ for excited states \cite{jacobs,bruneval} both place the $X_{5'}$ point near 5.5 eV although the gap to $X_3$ is wider than that found here. Inelastic effects in our calculation are considered to be insignificant because the energy is well below the onset of surface plasmons at 10.2 eV and bulk plasmons at 15.0 eV. However as a check we can include these effects from Im $\Sigma (E, \bm{k})$ in our calculation according to Eq.~(1) and Fig.~4 of Ref.~\onlinecite{read80}. We find that this inclusion has negligible effect on the energy position of this SR. The other barrier model with $z_0 = - 2.70$~\AA\ and $z_1 = -4.2$~\AA\ also produces the SR at this same energy of 5.3 eV. The surface band calculation of Ref.~\onlinecite{hummel} extending to 5 eV above $E_f$ also did not produce an SR at 4.0 eV at $\overline{M}/\overline{M}'$. 

\begin{figure}[b]
\includegraphics[scale=0.48]{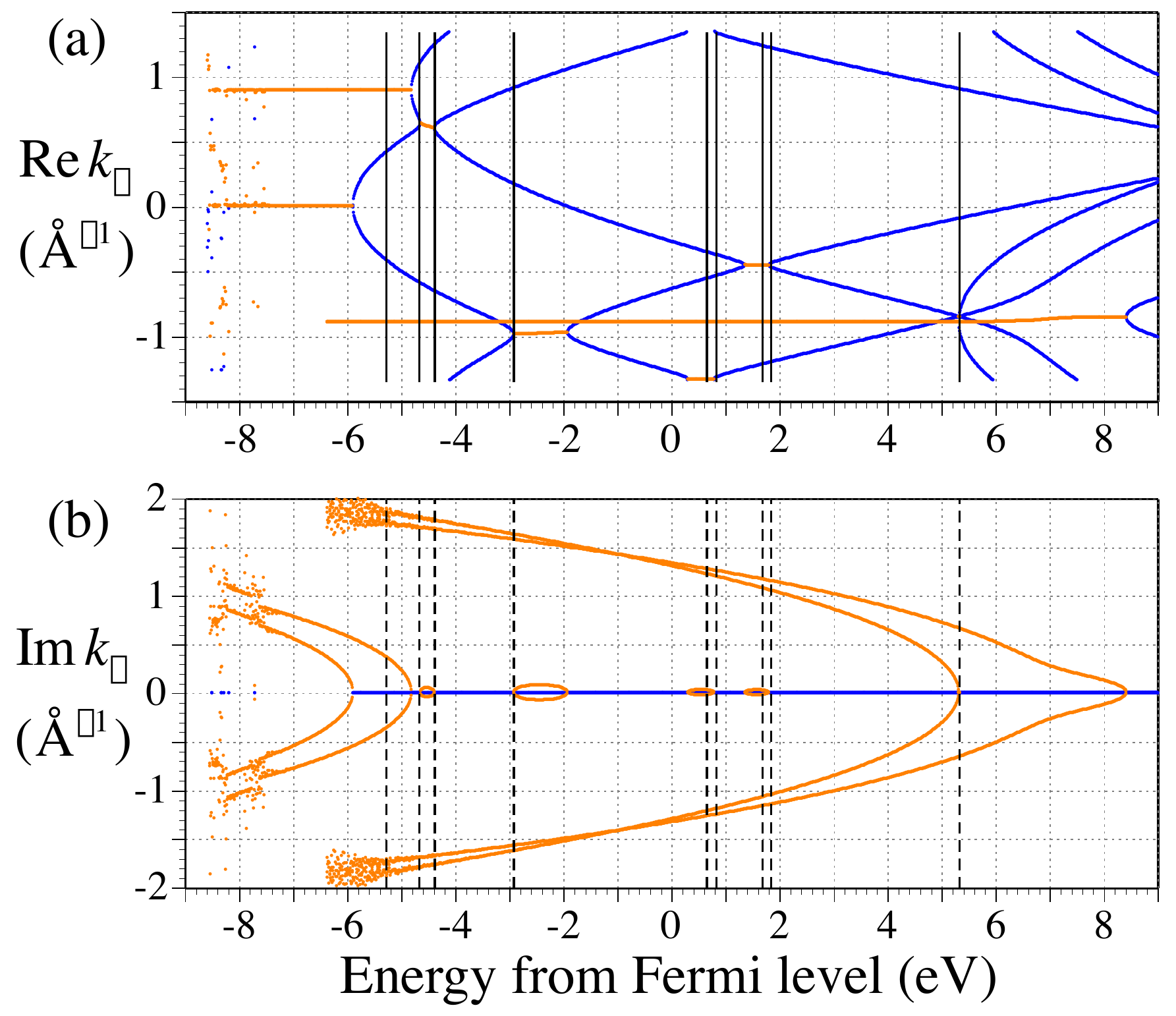}
\caption{(color online) The real and imaginary parts of the complex bulk and surface energy band structure $E(\bm{k}_{\parallel},k_{\perp})$ at $k_{\parallel} =1.21$~\AA$^{-1}$ along $\overline{\Gamma}$ -- $\overline{M}$ for Al(111) are shown. A very small gap occurs at 5.3 eV that is not discernable on the scale of the diagram.  Other features are the same as in Fig.~1(a) and (b).}
\end{figure}

\section{Calculation of surface bands for $\bm{\overline{\Gamma}}$ -- $\bm{\overline{M}}$.}

The above methods were used to calculate surface bands $E(\bm{k}_{\parallel})$ and the continuum of surface-projected bulk bands and their gaps for $\bm{k}_{\parallel}$ values corresponding to $\overline{\Gamma}$ -- $\overline{M}$ along the $\overline{\Sigma}$ direction and the results are plotted in Fig.~3. Approximately 40 values of $\bm{k}_{\parallel}$ were used to span this range of $\bm{k}_{\parallel}$. Thick full lines indicate surface states or strong resonances. Thinner full lines indicate weaker surface resonances. Resonances are classified as strong if in the present scattering picture they are formed where substrate reflection is strong. The dashed lines show the crystal empty-net or free-electron surface band structure that is also the $(E, \bm{k}_{\parallel})$ values at which plane waves in the region of constant potential between layers transition from evanescent to propagating in the crystal (i.e. acquire real $k_{\perp}$). To understand the dispersion and nature of the SS/SRs the full complex band structure in the near-surface region for a number of key values of $\bm{k}_{\parallel}$ for $\overline{\Gamma}$ -- $\overline{M}$ is examined.

Fig.~5 shows our full bulk and surface band structure for $k_{\parallel} = 1.21$~\AA$^{-1}$. Particularly notable is the gap near 1 eV at $\overline{M}$ in Fig.~4 becoming two partial gaps near 1.7 eV and 0.5 eV with SRs following inside these partial gaps. Note that there are SRs with small $\lambda_s$ also following just outside these partial gaps. These latter SRs are in the wide partial gaps formed from the higher energy band minima at 8.4 and 5.32 eV and $- \infty$. Below $E_f$ a new SR appears at $-5.28$ eV that is not a continuation of the SR at $-4.68$ eV. The SR at $-4.68$ eV and also the SR at $-4.39$ eV, have small $\lambda_s \approx 0.5$~\AA\ and are in the same wide gap mentioned above while the $-5.28$ eV SR is in the less-wide partial gap between the band minimum at $-4.8$ eV and $-\infty$ with $u_{\text{min}} \approx 0.38$~\AA$^{-1}$ and $\lambda_s \approx 2.6$~\AA.
 
Fig.~6 shows our full band calculation for $k_{\parallel} = 1.04$~\AA$^{-1}$ along $\overline{\Gamma}$ -- $\overline{M}'$. In their KRIPES study, Yang et al\cite{yang2} interpret their experimental feature at $\sim 4.2$ eV to be an SR while we calculate unoccupied SRs at 3.31, 3.43, 5.55 and 5.88 eV. The SRs at 3.43 and 5.88 eV have small values of $\lambda_s$ and are in the wide partial gap extending from the band minimum at 8.6 eV to $-\infty$. It is worth considering that interpretations from differences in the calculated bulk band structures could be involved. Reference~\onlinecite{yang2} shows their bulk band calculation for this $k_{\parallel}$ in their Fig.~6(a). They use an $l$-dependent self-consistent pseudo-potential method that is different from our method. However the two bulk band structures are nearly identical so that this is not a factor in the discrepancy. Note that there are image SRs near 8 eV also in the above wide partial gap with $\lambda_s$ values of $\approx 5.3$~\AA\ that are much larger than those at $\overline{\Gamma}$. Below $E_f$ the two lower energy SRs at $-2.37$ and $-1.29$ eV are also in this wide partial gap with small $\lambda_s$ while the SR at $-0.60$ eV is in a small width partial gap with $\lambda_s \approx 35$~\AA.

\begin{figure}
\includegraphics[scale=0.48]{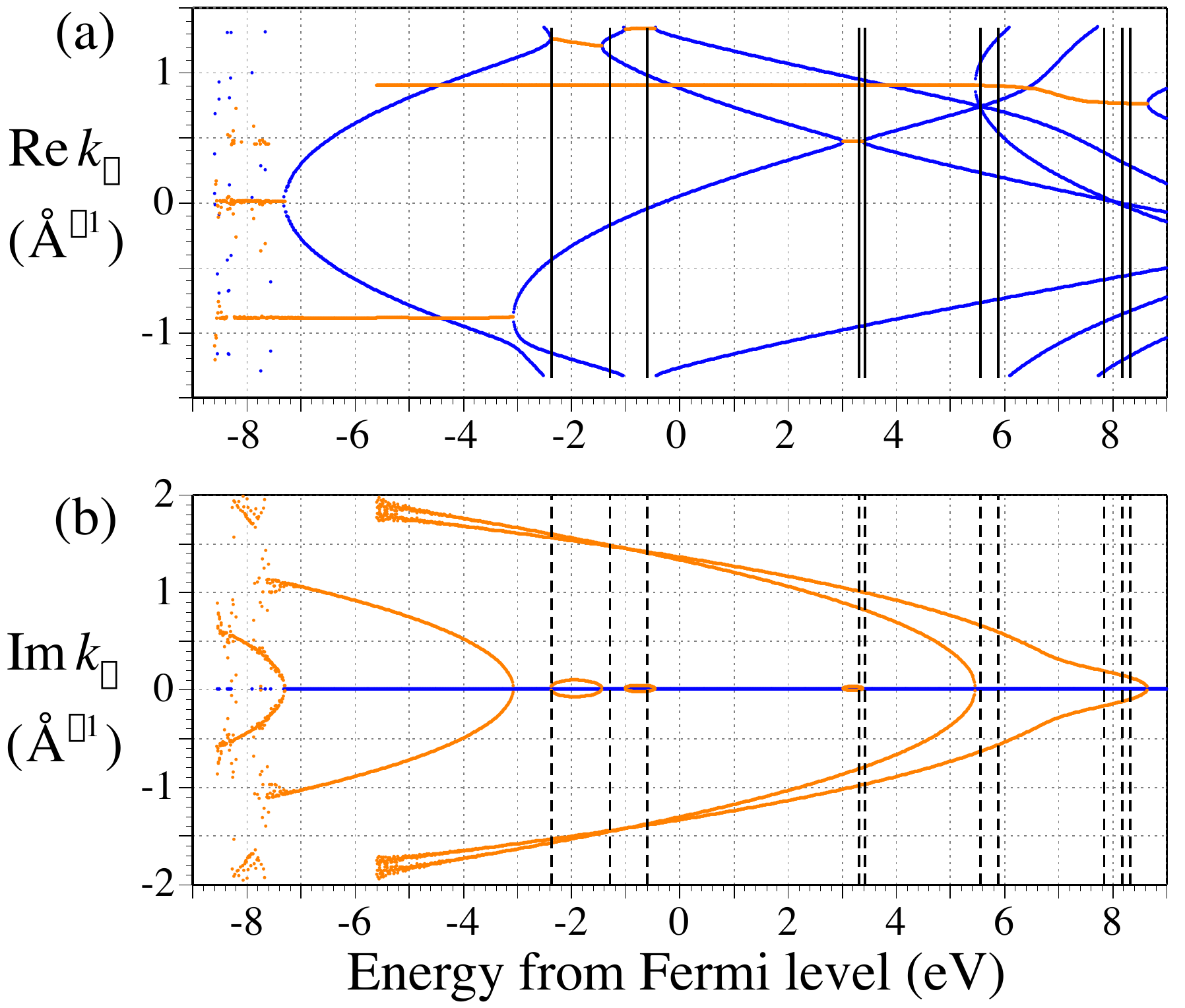}
\caption{(color online) The real and imaginary parts of the complex bulk and surface energy band structure $E(\bm{k}_{\parallel},k_{\perp})$ at $k_{\parallel} =1.04$~\AA$^{-1}$ along $\overline{\Gamma}$ -- $\overline{M}'$ for Al(111) are shown. A very small gap occurs at 5.5 eV that is not discernable on the scale of the diagram.  Other features are the same as in Fig.~1(a) and (b).}
\end{figure}

Fig.~7 shows our full band structure for $k_{\parallel} = 3/4 (\overline{\Gamma}\overline{M}) = 0.95$~\AA$^{-1}$. The bulk bands pass through the $K$ and $U$ points of the BBZ. The gap between the minima and maxima of the bands near 5.78 eV is still not visible on the scale of this diagram but the flatter odd symmetry band can be seen to have its minimum moved to 5.64 eV. There is an SR at 5.78 eV in the very small gap and an SR at 6.10 eV with $\lambda_s \approx 1.7$~\AA\ in the larger gap extending from the band minimum at 8.9 eV to $- \infty$. Two partial gaps of small width below $E_f$ in Fig.~6 now become one partial gap containing one SR at $-1.90$ eV just outside this partial gap with small $\lambda_s$ and one at $-0.86$ eV ($\lambda_s \approx 21$~\AA) inside it.

\begin{figure}
\includegraphics[scale=0.48]{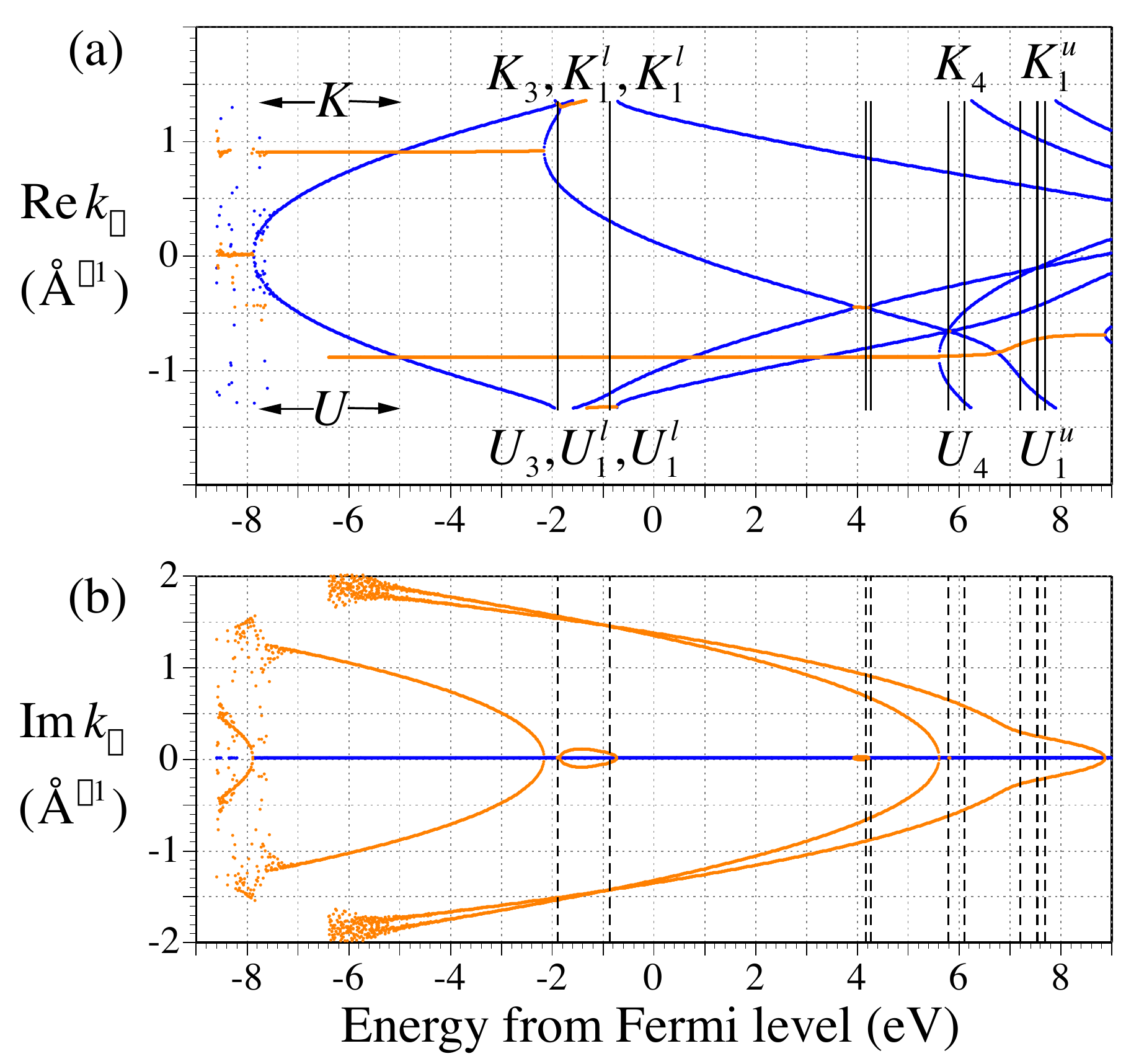}
\caption{(color online) The real and imaginary parts of the complex bulk and surface energy band structure $E(\bm{k}_{\parallel},k_{\perp})$ at $k_{\parallel} =0.95$~\AA$^{-1}\equiv 0.75 (\overline{\Gamma}\overline{M})$ along $\overline{\Gamma}$ -- $\overline{M}$  for Al(111) are shown. These $\bm{k}$ values pass along a line through the $K$ and $U$ points of the BBZ as indicated in panel (a). Bulk band labels at $K$ and $U$ are indicated at the margins. A very small gap occurs at 5.78 eV that is not discernable on the scale of the diagram.  Other features are the same as in Fig.~1(a) and (b).}
\end{figure}

In Fig.~8 for $k_{\parallel} = 0.92$~\AA$^{-1}$ a new very small partial gap appears at $-1.8$ eV that moves down in energy with decreasing $k_{\parallel}$ and persists to become the absolute $L$ gap at $\overline{\Gamma}$  in Fig.~1 and the SR from Fig.~7 at $-1.90$ eV continues as an SS in this case. The other partial gap with small width below $E_f$ has moved up in energy from its position in Fig.~7 along with the SR in it. At $k_{\parallel} < 0.75$~\AA$^{-1}$ for $\overline{\Gamma}$ -- $\overline{M}$ this partial gap disappears and so does the SR that has followed it as shown in Fig.~3. 

\begin{figure}
\includegraphics[scale=0.48]{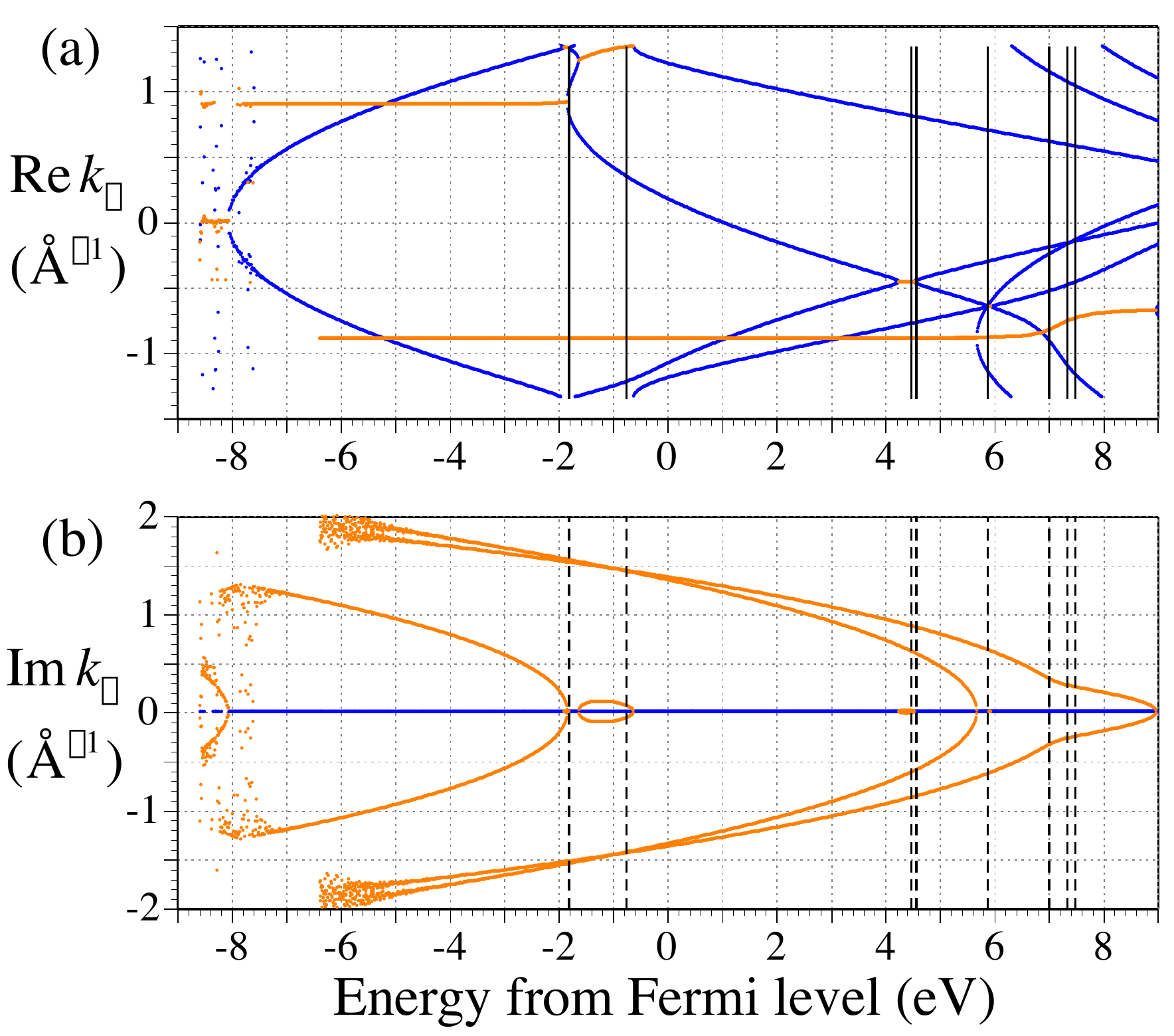}
\caption{(color online) The real and imaginary parts of the complex bulk and surface energy band structure $E(\bm{k}_{\parallel},k_{\perp})$ at $k_{\parallel} =0.95$~\AA$^{-1}$ along $\overline{\Gamma}$ -- $\overline{M}$ for Al(111) are shown.  Other features are the same as in Fig.~1(a) and (b).}
\end{figure}

There have been a number of studies of SS/SRs below $E_f$ for $\overline{\Gamma}$ -- $\overline{M}$ from ARPES. Kevan et al \cite{kevan} found the SS at $\overline{\Gamma}$  dispersed upwards and terminated at $-2.0$ eV for $k_{\parallel} \approx 0.9$~\AA$^{-1}$ and this agrees with our result shown in Fig.~3. Apparently they did not detect the SS continuation as an SR. Grepstad and Slagvold \cite{grepstad2} measured an SR dispersing from $-0.4$ eV at 1.08~\AA$^{-1}$ to $- 0.7$ eV at 0.95~\AA$^{-1}$ near 3/4($\overline{\Gamma}\overline{M}$) that agrees with our result. This SR dispersion was also measured by Hofmann and Kambe. \cite{hofman} In addition the latter find what they refer to as ``partial gap-edge features" from $-3$ to $-2$ eV at $k_{\parallel} = 1.15$~\AA$^{-1}$ to 0.90~\AA$^{-1}$ along $\overline{\Gamma}$ -- $\overline{M}'$ and also within this $\bm{k}_{\parallel}$ range near $-1.2$ eV. They do not detect these features along $\overline{\Gamma}$ -- $\overline{M}$ and hence do not consider them to be SRs. They consider them to be ``a kind of surface enhancement occurring at partial gap edges". However they coincide with our SRs in this range that have very small $\lambda_s \approx 0.7$~\AA\ and are in the wide partial gap from the higher energy band minima and $- \infty$ but just outside a smaller partial gap as shown in Figs.~6 and 7. These two SRs are very surface localized and appear to be difficult to detect in some theoretical methods and will be referred to later. 

\section{Calculation of surface states/resonances at $\bm{\overline{K}}$}

The calculated complex band structure for $E(\bm{k}_{\parallel}, k_{\perp}$) and $\bm{k}_{\parallel} =1.46$~\AA$^{-1} \equiv \overline{K}$ point is shown in Fig.~9(a) and (b) that also shows attenuating bulk states. Here there are five absolute projected gaps; two above and three below $E_f$ including the absolute gap below the band minimum at $-3.5$ eV. Real energy lines and half-loops extend from a band minima $\approx 15$ eV and produce a wide partial projected gap from this energy to $- \infty$. The result of the calculation for SS/SRs from Eq.~(\ref{eq;minimum}) is shown in Fig.~9(c) and energies plotted in Fig.~9(a) \& (b) and in Fig.~3 for $\overline{K}$ together with the absolute projected gaps. The minimum at $-0.45$ eV in Fig.~9(c) is not an SS of this system but indicates the crystal empty-net or surface free-electron energy as explained earlier. Above $-0.45$ eV only the three degenerate plane waves with reciprocal-net-vector coefficients $00$, $\overline{1}0$ and $0\overline{1}$ become propagating in this energy range. Below $-0.45$ eV these plane waves and all others are evanescent and have Re $k_{\perp}$ = 0 and finite Im $k_{\perp}$. The results in Fig.~9(c) determine that there are three SSs and two SRs below $E_f$ that are all type two in our classification. The SSs in the lowest narrow gap are at $-2.87$ and $-2.60$ eV with $\lambda_s$ of $\approx 23$~\AA\ and 53~\AA\ respectively. The SS at $-0.70$ eV has $\lambda_s \approx 11$~\AA. The SRs are not far outside the absolute gap edges with very small $\lambda_s \approx 0.4$~\AA\ in the very wide partial gap from $\approx 15$ eV to $-\infty$. Above  $-0.45$ eV the three degenerate plane waves labelled $00$, $\overline{1}0$ and $0\overline{1}$ have finite Re $k_{\perp}$ and are incident at points along the shape of the surface barrier. Above $E_f$ there are three SSs at 3.09, 3.13 and 3.20 eV with $\lambda_s \approx 31$, 33, 38~\AA\ respectively in one gap and two at 7.02 and 8.31 eV with $\lambda_s \approx 66$ and 16~\AA\ respectively in the higher energy gap. Note that there are also two SRs that are close to the absolute gap edges with very small $\lambda_s \approx 0.45$~\AA\ again in the partial gap from $\approx 15$ eV to $-\infty$.

\begin{figure}
\includegraphics[scale=0.48]{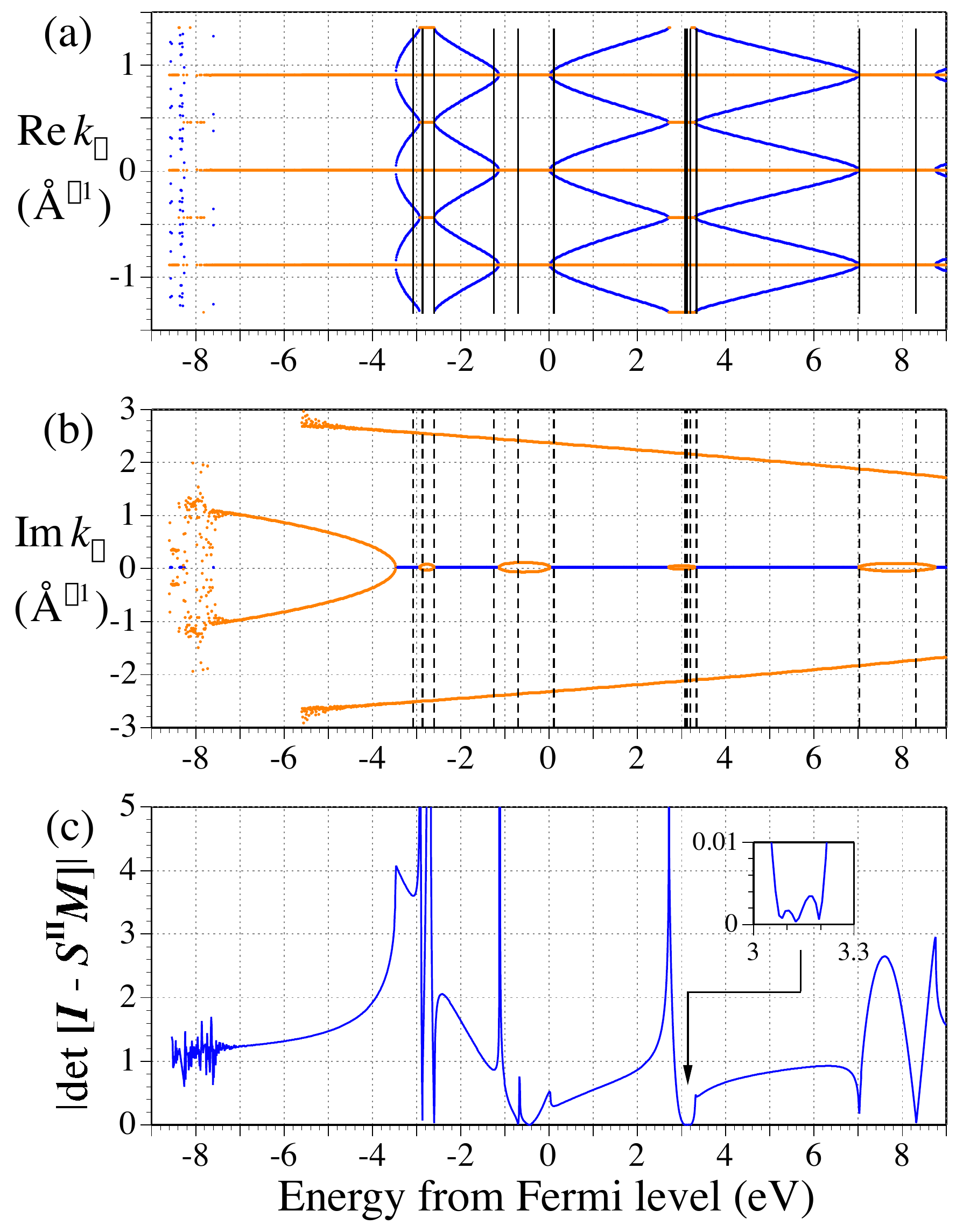}
\caption{(color online) The real and imaginary parts of the complex bulk and surface energy band structure $E(\bm{k}_{\parallel},k_{\perp})$ at $k_{\parallel} \equiv \overline{K}$ for Al(111) are shown in panels (a) and (b) respectively. Other features of panels (a) and (b) are the same as Fig.~1. Panel (c) shows the result of Eq.~(5) for the above $\bm{k}_{\parallel}$ value where minima indicate energies of surface states and resonances. The insert shows an expansion for the energy range 3 to 3.3 eV. The minium at $-0.45$ eV corresponds to a surface state for a free electron in the crystal.}
\end{figure}

Only the SS at $-0.7$ eV has been found from the ARPES experiments of Hofmann and Kambe \cite{hofman} and Kevan et al. \cite{kevan} 

\section{Calculations for $\bm{\overline{\Gamma}}$ -- $\bm{\overline{K}}$ -- $\bm{\overline{M}}$}

Approximately forty $\bm{k}_{\parallel}$ values corresponding to $\overline{\Gamma}$ -- $\overline{K}$ along the $\overline{T}$ direction were used to produce the surface band structure plotted in Fig.~3 from the calculated complex bulk band structures and using Eq.~(\ref{eq;minimum}). As this is not a bulk symmetry direction there are many SR bands degenerate with the bulk bands and the surface band structure is complicated. Kevan et al \cite{kevan} found the $- 0.7$ eV SS at $\overline{K}$ disperses upward towards $E_f$ for $\overline{K}$ -- $\overline{\Gamma}$ over a range of $k_{\parallel} \approx 0.3$~\AA$^{-1}$ and our calculated dispersion agrees with this. If one continues $\bm{k}_{\parallel}$ values along the $\overline{\Gamma}$ -- $\overline{K}$ direction then one arrives at $\overline{M}$ in the second SBZ. This was done to produce the surface band structures corresponding to $\overline{K}$ -- $\overline{M}$ along the $\overline{T}'$ direction that is plotted in Fig.~3 using approximately 40 values of $\bm{k}_{\parallel}$ and the same methods as before. All SS/SRs that have energy below the lowest energy surface free-electron band shown dashed in Fig.~3 are type two.

\section{Comparison with recent \textit{ab initio} surface barrier potential calculations}

The position of the geometric surface is defined as the plane where the bulk crystal is divided in half to create the semi-crystal. This is also the surface edge, $z_j$ for the jellium model of a semi-crystal. For Al(111) this plane extends a distance 2.21 a.u. (1.17~\AA) from the centre of the top row of atoms at $z=0$. The surface barrier height $U_0$ depends on the work function and the Fermi level, $E_f$ with respect to the constant potential energy in the surface atomic layer, $E_{c,1}$ or layer muffin-tin constant. Here $E_{c,1}$ is the same as the bulk constant potential, $E_c$. Both $E_c$ and $E_{c,1}$ have a contribution from the electron self-energy $\Sigma$ via the xc-potential in the crystal or atomic layer. Although we have used the experimental work function $\Phi$ of 4.24 eV, there is still some uncertainty in the bulk band structure calculation of $E_c$ and $E_f$. We have found that small differences in the values of $E_c$,   $E_f$ and $E_{c,1}$ and hence $U_0$ can significantly change the determination of the image plane position, $z_0$. Coupled with the measured uncertainty of $\pm 0.1$ eV in the first Rydberg image SR position at 0.46 eV from $E_v$ and no experimental measurement of the second image SR or position of the $L_{2'}$ and $L_1$ bands there is scope for some uncertainty in our value of $z_0$. As only Rydberg SS/SRs are very sensitive to $U_0$ and $z_0$ this uncertainty in $z_0$ has only a small effect on the energy position of other type-one SS/SRs. 

White et al \cite{white} have calculated the electronic self-energy $\Sigma$ experienced by electrons of the crystal in the near-surface region of Al(111) in a repeated-slab geometry of five Al layers and eight vacuum layers. They use the \textit{ab initio} GW method with the random-phase approximation for a dynamical screening interaction, W and a non-interacting approximation for the GreenÕs function, G from an initial density functional calculation with local-density approximation. Their non-local xc-potential gives the image potential and image-plane position, $z_0$. These authors find $|z_0| = 0.4 \pm 0.2$ a.u. from the geometric surface or $2.61\pm 0.2$ a.u. from $z = 0$ with saturation starting at $|z_1| \approx 6.0 +2.21 $ a.u. from $z = 0$. From their figures one determines that they found $U_0 \approx 9.5$ eV and $E_f \approx 4.8$ eV from $E_v$ and hence $\Phi \approx 4.8$ eV. In the following all SS/SRs are given at $\overline{\Gamma}$ with respect to $E_v$. They calculate an SS at 1.66 eV from $E_v$. Another repeated-slab calculation by Heinrichsmeier et al \cite{heinrich2} also using the GW approximation and a one-dimensional potential with several hundred layers found $|z_0| = 0.38$ a.u. from $z_j$ or 2.59 a.u. from $z =0$ and $\Phi = 4.82$ eV and found an image SR at $\approx 0.4$ eV. These two calculations of $\Phi$ have a difference of $\approx 0.6$ eV compared with the experimental value 4.24 eV.

Hence the present result of $|z_0| = 1.1$ a.u. (0.58~\AA) is less than half the GW \textit{ab initio} result of 2.6 a.u. (1.4~\AA) from $z = 0$ and inside the geometric surface in our case. This difference may be accounted for in part because of the differences in the barrier heights of these works compared with our value of $U_0 =12.84$ eV. If we set $U_0 = 9.5$ eV, $|z_0| = 2.61$ a.u. with $|z_1| = 8.21$ a.u. both from $z = 0$ in our calculation we calculate SRs at 1.7 eV and 0.2 eV. This is in agreement with the result in Ref.~\onlinecite{white}. With the same $z_0$ and $z_1$ values but with $U_0 = 12.84$ eV we calculate SRs at 2.6 eV and 0.3 eV. There is no experimental detection of SRs between $1.7$ and $2.6$ eV. There is discrepancy between $z_0$ (and $U_0$) from \textit{ab initio} and the present empirical model calculations. The value of $z_0$ found from either calculation cannot be considered definitive at this stage.  

\section{Conclusion}

In comparing the results of the present and three recent \cite{benesh,heinrich2,hummel} surface band calculations one is faced with explaining why these differ. All four agree on the occurrence of surface states in surface projected bulk gaps. The ABCM \cite{hummel} method finds only surface resonances that are the continuations of these surface states into partial projected bulk-gaps that have a small energy width and have real lines between bulk band minima and maxima. These resonances all have long decay lengths. The present scattering method produces resonance bands in close agreement with all of these bands but also produces additional bands. The slab DF \cite{heinrich2} and SEGF \cite{benesh} methods both do not produce these same resonance continuations of the states in all cases. The slab DF method does show broad resonances in some of these cases. Both the present scattering method and the slab DF calculation find a resonance band about 0.4 eV above the lower valence band edge along the $\overline{T}'$ direction ($\overline{M}$ -- $\overline{K}$) and extending also partially along $\overline{\Sigma}$ near $\overline{M}$ that is not found in either of the other calculations. In the present calculation this band and the one above it occur in the very wide partial bulk gap that contains attenuating bulk bands from real lines extending from a higher-energy bulk band minimum to $- \infty$. Both SRs are very close to the bulk band edges of a much smaller-width partial gap within this wider partial gap as shown at  $\overline{M}$ in Fig.~4. In our case of a continuum of bulk bands, the energies of these SRs do not also fall in the region of the smaller-width partial gap. However the finite layer slab DF method does not have a continuum of bulk bands and a small partial gap may have appeared in this energy region that may have elevated one of these bands to one with a long decay length and hence detectable in that method. 

It is suggested here that the above bands along $\overline{M}$ -- $\overline{K}$ are two members of a class of bands not detected in the ABCM method. Attenuating bulk states exist in very wide partial gaps where real lines extend from bulk band minima to $- \infty$ when there is no band maximum to which to join. We have found in the present work that SRs can form also from the real lines of these bulk band minima if their wave function matches that through the surface potential transition region into the vacuum just as for other slower attenuating bulk states from small-width gaps. The lower the energy of the SR from the bulk band minimum the shorter the decay length provided the SR energy does not also lie in a smaller-width absolute or partial gap. An SR in this type of wide gap does not always have a very short decay length and an example is shown for $k_{\parallel}=1.04$~\AA$^{-1}$ for $\overline{\Gamma}$ -- $\overline{M}'$ in Fig.~6 for the image SRs near 8 eV. Where these SRs have very short decay lengths their effect on photoemission intensity profiles and whether they can be detected experimentally remains to be determined. In the cases where these SRs are found theoretically to lie also just outside the edges of small energy-width partial bulk gaps it is possible that a small change to the bulk or surface potentials could move them inside this gap and change the decay length from short to long.

It is also suggested here that unexplained features in experimental photoemission previously described as surface enhancement occurring at partial bulk gap-edges or enhancement of bulk density-of-states near the surface are actually SRs of the above type. This was discussed in Sec.~VII. It is important for theoretical surface band structure methods to be capable of identifying all SRs including this type which may be very localized to the surface atomic layer and if type two (scattering from the bottom of the surface barrier potential) may not extend far into the vacuum either. Rydberg image resonances may also have very short decay lengths in the crystal as in the case here at $\overline{\Gamma}$  but they are type one (scattering near the top of the surface barrier) and they extend far into the vacuum. 

The present scattering method may have some advantages over other surface band structure methods. Sufficient accuracy is obtained before the onset of numerical instabilities because of the rapid convergence obtained in the layer scattering calculation from direct and reciprocal net summations. The locations of SS/SR bands depend strongly on the accuracy of the bulk potential. In the present method a bulk potential is chosen \textit{a priori} from a bulk band structure that has verified accuracy from comparison with experiment. The empirically constructed surface potential has proven to be satisfactory since it reproduces all of the surface band structure obtained in the \textit{ab initio} self-consistent warped muffin-tin surface potential used in the ABCM method. The scattering method identifies many SRs that have very short decay lengths and are mainly localized within the surface layer. These SRs have not been detected in the ABCM or other calculations. In the scattering method these SRs are just as easily detected as the SRs of long decay length. In addition it produces the image SS/SRs and it can be extended to unbound SRs in higher energy regions where inelastic effects become significant. The method also provides a single picture for understanding the formation of what have historically been termed Shockley, Tamm and Rydberg SS/SRs. It therefore provides a complete surface band structure that can aid in the analysis of experimental features from band transitions in photoemission and other surface spectroscopies. The use of the method for other metallic surfaces and those with adlayers of foreign atoms is to be determined.

\end{document}